\documentclass[conference]{IEEEtran}
\IEEEoverridecommandlockouts

\usepackage{cite}
\usepackage{amsmath,amssymb,amsfonts}
\usepackage{algorithmic}
\usepackage{graphicx}
\usepackage{textcomp}
\usepackage{xcolor}
\usepackage{glossaries-extra}
\usepackage{siunitx}
\usepackage{booktabs}
\usepackage{tikz}
\usepackage{hyperref}
\usetikzlibrary{arrows.meta}
\usepackage{comment}
\usepackage{pgfplots}
\pgfplotsset{compat=1.18} 
\def\BibTeX{{\rm B\kern-.05em{\sc i\kern-.025em b}\kern-.08em
    T\kern-.1667em\lower.7ex\hbox{E}\kern-.125emX}}

\makeatletter
\newcommand{\linebreakand}{%
  \end{@IEEEauthorhalign}
  \hfill\mbox{}\par
  \mbox{}\hfill\begin{@IEEEauthorhalign}
}
\makeatother

\begin{document}
\title{High-Resolution Range-Doppler Imaging from One-Bit PMCW Radar via Generative Adversarial Networks}
\newabbreviation{ML}{ML}{machine learning}
\newabbreviation{PMCW}{PMCW}{phase-modulated continuous wave}
\newabbreviation{FMCW}{FMCW}{frequency-modulated continuous wave}
\newabbreviation{ADAS}{ADAS}{advanced driver assistance system}
\newabbreviation{ADC}{ADC}{analog-to-digital converter}
\newabbreviation{RD}{RD}{range-Doppler}
\newabbreviation{CDM}{CDM}{code-division multiplexing}
\newabbreviation{RF}{RF}{radio frequency}
\newabbreviation{OFDM}{OFDM}{orthogonal frequency-division multiplexing}
\newabbreviation{CPI}{CPI}{coherent processing interval}
\newabbreviation{ECB}{ECB}{equivalent complex baseband}
\newabbreviation{RCS}{RCS}{radar cross-section}
\newabbreviation{DFT}{DFT}{discrete Fourier transform}
\newabbreviation{E2E}{E2E}{end-to-end}
\newabbreviation{NN}{NN}{neural network}
\newabbreviation{SNR}{SNR}{signal-to-noise ratio}
\newabbreviation{PSNR}{PSNR}{peak signal-to-noise ratio}
\newabbreviation{SSIM}{SSIM}{structural similarity index}
\newabbreviation{PSR}{PSR}{peak-to-sidelobe ratio}
\newabbreviation{ISL}{ISL}{integrated sidelobe level}
\newabbreviation{FFT}{FFT}{fast Fourier transform}
\newabbreviation{GAN}{GAN}{generative adversarial network}
\newabbreviation{PN}{PN}{pseudo-noise}
\newabbreviation{MLS}{MLS}{maximum length sequence}
\newabbreviation{AWGN}{AWGN}{additive white Gaussian noise}
\newabbreviation{GP}{GP}{gradient penalty}
\newabbreviation{MSE}{MSE}{mean squared error}
\newabbreviation{AD}{AD}{automated driving}
\newabbreviation{UAV}{UAV}{unmanned aerial vehicle}
\newabbreviation{MIMO}{MIMO}{multiple-input multiple-output}
\newabbreviation{CNN}{CNN}{convolutional neural network}
\newabbreviation{GT}{GT}{ground truth}
\newabbreviation{SISO}{SISO}{single-input single-output}
\newabbreviation{HR}{HR}{high-resolution}
\newabbreviation{ReLU}{ReLU}{rectified linear unit}
\newabbreviation{DNN}{DNN}{deep neural network}
\newabbreviation{BN}{BN}{batch normalization}
\newabbreviation{Conv}{Conv}{convolution}
\newabbreviation{FLOPS}{FLOPS}{floating-point operations per second}
\newabbreviation{ResNet}{ResNet}{residual neural network}
\newabbreviation{PSL}{PSL}{peak-sidelobe level}
\author{\IEEEauthorblockN{Jingxian Wang}
\IEEEauthorblockA{
\textit{FAU Erlangen-Nürnberg}\\
Erlangen, Germany \\
jingxian.wang@fau.de}
\and
\IEEEauthorblockN{Moritz Kahlert}
\IEEEauthorblockA{
\textit{HELLA GmbH \& Co. KGaA}\\
Lippstadt, Germany \\
moritz.kahlert@forvia.com}
\and
\IEEEauthorblockN{Tai Fei}
\IEEEauthorblockA{
\textit{University of Applied Sciences and Arts}\\
Dortmund, Germany \\
tai.fei@fh-dortmund.de}
\linebreakand
\IEEEauthorblockN{Changxu Zhang}
\IEEEauthorblockA{
\textit{HELLA GmbH \& Co. KGaA}\\
Lippstadt, Germany \\
changxu.zhang@forvia.com}
\and
\IEEEauthorblockN{Zhaoze Wang}
\IEEEauthorblockA{
\textit{HELLA GmbH \& Co. KGaA}\\
Lippstadt, Germany \\
zhaoze.wang@forvia.com}
\and
\IEEEauthorblockN{Markus Gardill}
\IEEEauthorblockA{
\textit{Brandenburg University of Technology}\\
Cottbus, Germany \\
markus.gardill@b-tu.de}
}

\maketitle
\begin{abstract}
Digital modulation schemes such as \gls{PMCW} have recently attracted increasing attention as possible replacements for \gls{FMCW} modulation in future automotive radar systems. 
A significant obstacle to their widespread adoption is the expensive and power-consuming \glspl{ADC} required at gigahertz frequencies. 
To mitigate these challenges, employing low-resolution \glspl{ADC}, such as one-bit, has been suggested.
Nonetheless, using one-bit sampling results in the loss of essential information.
This study explores two \gls{RD} imaging methods in \gls{PMCW} radar systems utilizing \glspl{NN}. 
The first method merges standard \gls{RD} signal processing with a \gls{GAN}, whereas the second method uses an \gls{E2E} strategy in which traditional signal processing is substituted with an \gls{NN}-based \gls{RD} module.
The findings indicate that these methods can substantially improve the probability of detecting targets in the range-Doppler domain.
\end{abstract}

\glsresetall
\section{Introduction}
\label{sec:introduction}

Digital modulation schemes, such as \gls{PMCW}, have recently attracted attention due to their robustness against mutual interference and their inherent multiplexing capability, which is essential for \gls{MIMO} systems \cite{9318758}. However, \gls{PMCW} and other digital modulation schemes require fast-sampling \glspl{ADC} to process the entire baseband, reaching up to \SI{1}{\giga\hertz} at \SI{77}{\giga\hertz} and \SI{4}{\giga\hertz} at \SI{79}{\giga\hertz}.
These high-sampling \glspl{ADC} are power-consuming and generate significantly larger data volumes than \gls{FMCW} radar systems, which only need to sample a narrowband beat signal \cite{4404963}.
To address this issue, with a focus on waveform design and signal processing, stepped frequency solutions have been proposed for \gls{PMCW} \cite{10838615}, providing a tradeoff between performance and hardware requirements.
The carrier frequency changes linearly over time to reduce the bandwidth of each pulse, which in turn lowers the \gls{ADC} sampling requirements.
An alternative approach to reducing sampling rates involves lowering the resolution of the \glspl{ADC} \cite{9695370} and, for example, employing them in mixed-\gls{ADC} setups, as shown in \cite{10636418}.
Low-resolution \glspl{ADC}, e.g., one-bit \glspl{ADC}, could be a promising solution to reduce the amount of data, power consumption, and costs.
However, amplitude information is lost by one-bit sampling, which provides crucial information for target detection.
In \cite{10648744}, the authors focused on designing the transmit code and receive filter in the presence of one-bit sampling.

This study investigates the effectiveness of \gls{NN}-based methods, specifically \gls{E2E} and hybrid approaches, to improve target detectability when applying one-bit quantization.
Our contributions include the development of an \gls{E2E} approach for \gls{RD} map generation based on high-resolution and one-bit quantized \gls{ADC} data and the introduction of a denoising network to mitigate noise introduced by quantization. 
Furthermore, we propose a hybrid approach combining conventional \gls{RD} processing with a denoising network. 
Finally, we provide a comparative performance evaluation of the \gls{E2E} and hybrid approaches, highlighting their advantages.

\section{Signal Model}
\label{sec:signal_model}
Let $\mathbf{x} = [x_0, \dots, x_{N-1}] = \exp(\jmath \boldsymbol{\phi})$ with $\boldsymbol{\phi} \in \{0, \pi\}^{N}$ represent a \gls{PN} binary sequence of length $N$, where $x_n \in \{-1, 1\}$ is denoted as a chip.
The transmitted and received signals in their \gls{ECB} representation can be expressed by $x_\mathrm{BB}$ and $y_\mathrm{BB}$, respectively,
\begin{equation}
    x_\mathrm{BB}(t) = \sum\limits_{n=0}^{N-1}x_{n}\mathrm{rect}\left(\frac{t-(n+0.5)T}{T}\right),
\end{equation}
and
\begin{equation}
        y_\mathrm{BB}(t) = \sum\limits_{k=0}^{K-1} \gamma_k \,x_\mathrm{BB} \bigl(t-\tau_k(t)\bigr)\exp\bigl(-\jmath 2 \pi f_\mathrm{c} \tau_k(t)\bigr),
\end{equation}
where $T$ is the chip duration, $\mathrm{rect}(\cdot)$ denotes the rectangular function, $K$ is the number of point targets, $\gamma_k$ is the scaling factor of the $k$th target reflection, including free-space attenuation, \gls{RCS}, and the reflection phase, $\tau_k(t)$ is the round-trip delay, and $f_\mathrm{c}$ is the carrier frequency. 
The round-trip delay can be expressed by $\tau_k(t) = 2r_{0, k}/c_0 + 2v_{\mathrm{r},k}t/c_0$, where $r_{0,k}$ is the range between the radar and the $k$th target at the beginning of the \gls{CPI}, $v_{\mathrm{r}, k}$ is the relative velocity, and $c_0$ is the light speed.
In addition, it is assumed that $M$ sequences with sequence duration $T_\mathrm{seq}$ are transmitted in a single \gls{CPI}, and $y_\mathrm{BB}(t)$ is sampled with a rate of $1/T$ at time steps $t_\mathrm{s} = nT+mT_\mathrm{seq}$ with $0\leq n <N, 0\leq m <M$, 
resulting in the two-dimensional matrix representation of the sampled baseband signal expressed by $\mathbf{Y} = y_\mathrm{BB}(t_\mathrm{s}) \in \mathbb{C}^{N \times M}$.

Further, it is assumed that the output of the \glspl{ADC} can be either high-resolution or one-bit. 
The output of the \glspl{ADC} after one-bit quantization can be expressed by $\tilde{\mathbf{Y}} = \mathcal{Q}(\mathbf{Y}) = \mathrm{sign}\big(\Re(\mathbf{Y})\big) + \jmath\, \mathrm{sign}\big(\Im(\mathbf{Y})\big)$, where $\mathcal{Q}(\cdot)$ denotes the complex quantization operator, $\mathrm{sign}(x)$ is the sign function, and $\Re(\cdot)$ and $\Im(\cdot)$ denote the real and imaginary parts, respectively.
We define that $\mathrm{sign}(x) = 1$ for $x \ge 0$, otherwise $\mathrm{sign}(x) = -1$.
 
To retrieve the range and Doppler information from the baseband signal, we first apply cross-correlation along the fast-time domain of $\mathbf{Y}$ as follows, 
\begin{equation}
    p_{rm} = \sum\limits_{n=0}^{N-1}x^*_{\bmod\left(n-r, N\right)} y_{nm},
    \label{eq:range_profile}
\end{equation}
where $(y_{nm})$ is the $(r, m)$th element in $\mathbf{Y}$, $\mathbf{P}= (p_{rm}) \in \mathbb{C}^{N\times M}$ is the range profile, $r$ denotes the range bin index, $(\cdot)^*$ is the complex conjugate, and $\bmod(\cdot)$ is the modulo operator.
Subsequently, the relative velocities (i.e., Doppler shifts)  can be extracted by applying \glspl{DFT} along the slow-time domain, 
resulting in the \gls{RD} map $\mathbf{Q} = (q_{rv}) \in \mathbb{C}^{N\times M}$. The $(r, v)$th element can be expressed by
\begin{equation}
    q_{rv} = \sum\limits_{m=0}^{M-1}p_{rm}\exp\left(-\jmath 2\pi v \frac{m}{M} \right).
    \label{eq:DopplerProcessing}
\end{equation}
Note that by replacing $\mathbf{Y}$ with $\tilde{\mathbf{Y}} = (\tilde{y}_{nm})$ in \eqref{eq:range_profile}, $\tilde{\mathbf{R}}$ and $\tilde{\mathbf{S}}$ can be calculated similarly without losing generality. 
\section{Range-Doppler Neural Networks}

This work presents two \gls{NN}-based approaches for generating and denoising \gls{RD} maps from 1-bit \gls{ADC} data: an \gls{E2E} approach and a hybrid approach. 
Both networks are adversarially trained within a \gls{GAN} framework. 
The \gls{E2E} method integrates \glspl{CNN} \cite{726791}, \glspl{ResNet} \cite{7780459}, and frequency domain operations.
The hybrid approach combines conventional signal processing techniques with a denoising \gls{NN} to improve performance.

\subsection{Network Architecture}
\begin{figure}
  \centering
  \usetikzlibrary{shapes.geometric, arrows}

\definecolor{mycolor1}{rgb}{0.00000,0.44700,0.74100}%
\definecolor{mycolor2}{rgb}{0.85000,0.32500,0.09800}%
\definecolor{mycolor3}{rgb}{0.92900,0.69400,0.12500}%
\definecolor{mycolor4}{rgb}{0.49400,0.18400,0.55600}%
\definecolor{mycolor5}{rgb}{0.46600,0.67400,0.18800}%
\definecolor{mycolor6}{rgb}{0.30100,0.74500,0.93300}%
\definecolor{mycolor7}{rgb}{0.63500,0.07800,0.18400}%

\tikzstyle{startstop} = [rectangle, rounded corners, minimum width=2cm, minimum height=1cm,text centered, draw=black]
\tikzstyle{process} = [rectangle, minimum width=2cm, minimum height=1cm, text centered, draw=black]
\tikzstyle{data} = [rectangle, minimum width=3cm, minimum height=1cm, text centered, draw=black]
\tikzstyle{output} = [rectangle, minimum width=3cm, minimum height=1cm, text centered, draw=black]
\tikzstyle{decision} = [rectangle, minimum width=3cm, minimum height=1cm, text centered, draw=black]
\tikzstyle{result} = [ellipse, minimum width=3cm, minimum height=1cm, text centered, draw=black]
\tikzstyle{arrow} = [thick,-Stealth,>=stealth]

\begin{tikzpicture}[node distance=2cm]


    \node (input) [startstop, align=center] {1-bit\\\glsxtrshort{ADC} data};
    \node (rd) [rectangle, rounded corners, minimum width=2.0cm, minimum height=1cm,text centered, draw=black, below of=input, align=center, yshift=0.5cm]{\glsxtrshort{RD}\\ processing};
    \node (noise) [rectangle, rounded corners, minimum width=2.0cm, minimum height=1cm,text centered, draw=black, below of=rd, anchor=north, yshift=1cm, align=center]{Denoising};
    \node (gen) [rectangle, rounded corners, minimum width=2.8cm, minimum height=2.5cm,text centered, draw=black, align=center, below of=input, anchor=north, yshift=1cm] {};
    \node (gt) [startstop, right of=noise, xshift=4cm, align=center] {\Glsxtrlong{HR} \\ 
    \glsxtrshort{RD} map};
    \node (disc) [startstop, below of=noise, xshift=3cm, align=center, yshift=0.5cm] {Discriminator\\(PatchGAN)};
    \node (auth) [startstop, below of=disc, align=center, yshift=0.3cm] {Authenticity \\score};
    \node (gentext)[left of=gen, rotate=90, yshift=-0.8cm]{Generator};

    \draw [arrow] (input) -- (rd);
    \draw [arrow] (rd) -- (noise);
    \draw [arrow] (noise) |- (disc);
    \draw [arrow] (gt) |- (disc);
    \draw [arrow] (disc) -- (auth);
    \draw [dashed,-Stealth,>=stealth] (auth.south) -- ++ (0, -0.5)
    coordinate (A) -- ++(-4.7,0) node [above, midway] {Adversarial feedback}
    coordinate (B) -- ++(0,4.95) 
    -- (gen.west);

\end{tikzpicture}
  \caption{Overall \glsxtrlong{NN} architecture for \glsxtrlong{RD} map generation and denoising based on 1-bit \glsxtrshort{ADC} data.}
  \label{fig:network_architecture}
\end{figure}
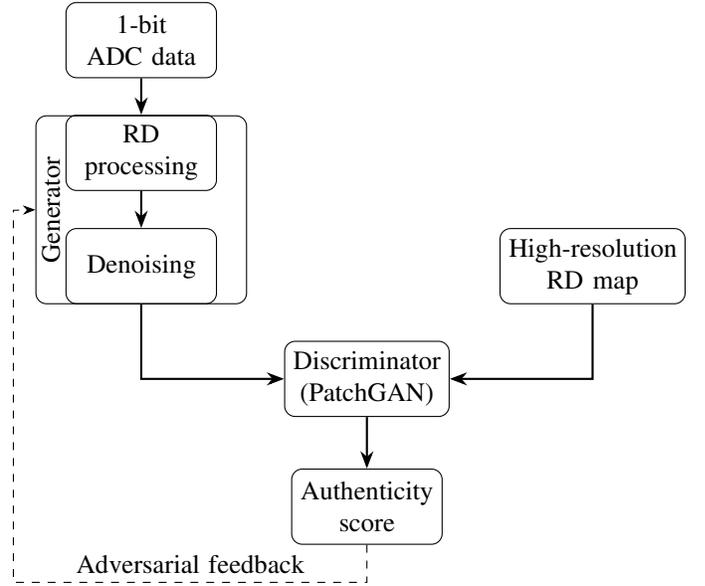

In our \gls{GAN} framework (Fig.~\ref{fig:network_architecture}), the generator processes 1-bit \gls{ADC} data to reconstruct high-quality \gls{RD} maps, while the discriminator, implemented as a PatchGAN, assesses their authenticity by comparing them to \gls{HR} \gls{RD} maps from a full-precision \gls{ADC}. The \gls{HR} \gls{RD} map serves as a reference, enabling the discriminator to refine the generator iteratively by providing authenticity feedback, thereby improving the fidelity of the generated outputs. 

In the \gls{E2E} approach, all generator layers are trainable.
In contrast, in the hybrid approach, the initial layers are replaced with classical \gls{RD} processing, as described in \eqref{eq:range_profile} and \eqref{eq:DopplerProcessing}, while the remaining layers, i.e., the backbone network, focus on noise reduction.

\subsection{Generator in E2E Approach}
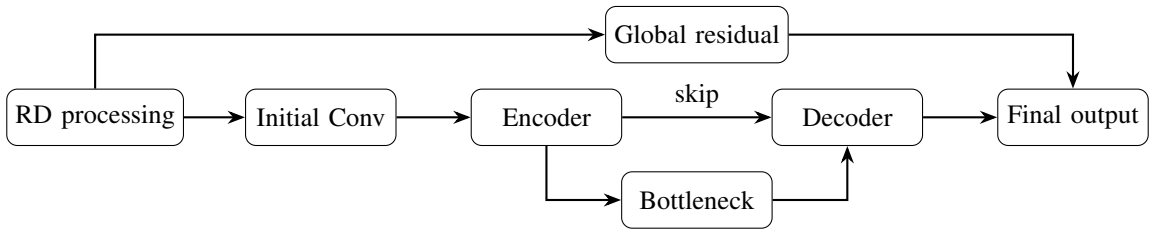
\begin{figure*}
    \centering
    \usetikzlibrary{shapes.geometric, arrows}

\definecolor{mycolor1}{rgb}{0.00000,0.44700,0.74100}%
\definecolor{mycolor2}{rgb}{0.85000,0.32500,0.09800}%
\definecolor{mycolor3}{rgb}{0.92900,0.69400,0.12500}%
\definecolor{mycolor4}{rgb}{0.49400,0.18400,0.55600}%
\definecolor{mycolor5}{rgb}{0.46600,0.67400,0.18800}%
\definecolor{mycolor6}{rgb}{0.30100,0.74500,0.93300}%
\definecolor{mycolor7}{rgb}{0.63500,0.07800,0.18400}%

\tikzstyle{startstop} = [rectangle, rounded corners, minimum width=2cm, minimum height=0.75cm,text centered, draw=black]
\tikzstyle{data} = [rectangle, minimum width=3cm, minimum height=0.75cm, text centered, draw=black]
\tikzstyle{output} = [rectangle, minimum width=3cm, minimum height=0.75cm, text centered, draw=black]
\tikzstyle{decision} = [rectangle, minimum width=3cm, minimum height=0.75cm, text centered, draw=black]
\tikzstyle{result} = [ellipse, minimum width=3cm, minimum height=0.75cm, text centered, draw=black]
\tikzstyle{arrow} = [thick,-Stealth,>=stealth]

\begin{tikzpicture}[node distance=2cm]

    \node (pmcw) [startstop, align=center] {\gls{RD} processing};
    \node (init) [startstop, align=center, right of=pmcw, xshift=1cm] {Initial Conv};
    \node (encoder) [startstop, align=center, right of=init, xshift=1cm] {Encoder};
    \node (decoder) [startstop, align=center, right of=encoder, xshift=2cm] {Decoder};
    \node (final) [startstop, align=center, right of=decoder, xshift=1cm] {Final output};
    \node (bottleneck) [startstop, align=center, below of=encoder, xshift=2cm, align=center, yshift=0.9cm] {Bottleneck};
    \node (residual) [startstop, align=center, above of=encoder, align=center, xshift=2cm, yshift=-0.9cm] {Global residual};

    \draw [arrow] (pmcw) -- (init);
    \draw [arrow] (init) -- (encoder);
    \draw [arrow] (encoder) -- (decoder) node [above, midway] {skip};
    \draw [arrow] (decoder) -- (final);
    \draw [arrow] (pmcw) |- (residual);
    \draw [arrow] (residual) -| (final);
    \draw [arrow] (encoder) |- (bottleneck);
    \draw [arrow] (bottleneck) -| (decoder);

\end{tikzpicture}
    \caption{Structure of the proposed generator network. The functionalities of the blocks are outlined in Table~\ref{tab:generator_structure}.}
    \label{fig:generator_diagram}
\end{figure*}

To design the generator in the \gls{E2E} approach, we adopt the idea from \cite{yang2023adcnetlearningrawradar} and integrate a similar learnable signal processing layer into the backbone network, which is the Pix2Pix generator from \cite{8100115}, as illustrated in Fig.~\ref{fig:generator_diagram}. This modified Pix2Pix generator is responsible for both \gls{RD} processing and noise reduction to reconstruct high-quality \gls{RD} maps.

The learnable signal processing layer in \cite{yang2023adcnetlearningrawradar} was originally designed for \gls{FMCW} radar, where range estimation is performed through \gls{DFT}. 
However, in \gls{PMCW} radar, the range profile is obtained by correlation, as described in \eqref{eq:range_profile}. To accommodate this difference, we replace the \gls{DFT}-based range estimation with a frequency-domain correlation operation. Specifically, we employ a fixed-length, trainable \gls{PN} sequence as the correlation kernel, which is optimized during training to improve the correlation response to target echoes, thereby improving the accuracy of the range profile.
Unlike the hybrid approach, where the \gls{PN} sequence is predefined and fixed \textit{a priori}, the \gls{E2E} approach enables data-driven learning of the \gls{PN} sequence. 
Additionally, analogous to \eqref{eq:DopplerProcessing}, Doppler processing is implicitly learned within the network. 
These modifications improve both accuracy and robustness in the \gls{RD} processing.

The Pix2Pix generator, originally based on a U-Net \cite{10.1007/978-3-319-24574-4_28} architecture, employs skip connections to transfer low-level spatial information from the encoder to the decoder. 
However, applying Pix2Pix directly to 1-bit \gls{ADC} data leads to training instability and suboptimal reconstruction, as radar signals exhibit characteristics distinct from natural images. 
To mitigate these issues, we introduce three key modifications. 
First, the network depth is increased with a bottleneck layer to enhance feature extraction and stabilize training, improving gradient flow and preventing vanishing gradients. 
Second, residual learning is incorporated, where, instead of relying solely on U-Net’s skip connections, multiple residual blocks \cite{7780459} are integrated within both the encoder and decoder, improving gradient propagation and ensuring stable training. 
Third, a global residual connection is introduced using a $1 \times 1$ \gls{Conv}, directly linking the input to the output to preserve low-frequency components in the final \gls{RD} map. These enhancements improve stability, robustness, and accuracy, enabling effective training on 1-bit \gls{ADC} data and high-fidelity \gls{RD} map reconstruction.

\begin{table}
    \centering
    \caption{Generator architecture and its components functions.}
    \begin{tabular}{p{0.2\columnwidth}p{0.35\columnwidth}p{0.32\columnwidth}}
        \toprule
        \textbf{Component} & \textbf{Layer Type} & \textbf{Main Function} \\
        \midrule
        \glsxtrshort{RD} processing & Correlations + \glspl{DFT} & Extracts range and Doppler information \\
        Initial \glsxtrshort{Conv} & $3 \times 3$ \glsxtrshort{Conv} + \glsxtrshort{ReLU} & Feature extraction \\
        Encoder & $4 \times 4$ \glsxtrshort{Conv} + \gls{BN} + \glsxtrshort{ReLU} + residual block (×4) & Downsampling and feature enhancement \\
        Residual block &  $3 \times 3$ \glsxtrshort{Conv} + \gls{BN} + \glsxtrshort{ReLU} + skip connection & Mitigates vanishing gradient and enables deeper network training \\
        Bottleneck & 3 residual blocks & Deep feature transformation \\
        Decoder & $4 \times 4$ transposed \glsxtrshort{Conv} + \glsxtrshort{BN} + \glsxtrshort{ReLU} + residual block (×4) & Upsampling path for reconstruction \\
        Global residual & $1 \times 1$ \glsxtrshort{Conv} & Directly connects input to output \\
        Final output & $3 \times 3$ \glsxtrshort{Conv} + Tanh & Generates final \gls{RD} map \\
        \bottomrule
    \end{tabular}
    \label{tab:generator_structure}
\end{table}

\subsection{Discriminator}
The discriminator distinguishes between reference, specifically \gls{HR}, and generated \gls{RD} maps, providing adversarial feedback to the generator and motivating it to produce more authentic output.
The design utilizes a PatchGAN discriminator \cite{8100115}, focusing on classifying localized regions instead of whole images. This approach enables the discriminator to discern intricate details.
The components of the discriminator are outlined as follows.
Firstly, feature extraction is performed using three $4 \times 4$ \gls{Conv} layers with a stride of 2 for downsampling. 
Each layer is succeeded by Leaky \gls{ReLU} activation to derive hierarchical features from the input data.
Secondly, PatchGAN employs a patch-based discrimination approach that evaluates smaller sections rather than making an overarching decision on the full \gls{RD} map, thereby improving the retention of detailed features.
In the third and last discriminative stage, the concluding pair of layers utilizes fully connected \glspl{Conv} to generate a singular scalar output representing the probability of an authentic input \gls{RD} map.

Using a PatchGAN enhances the ability of the discriminator to help the generator preserve intricate structural information, thereby improving the reconstruction performance.
The structure of the discriminator network is outlined in Table~\ref{tab:discriminator_structure}.

\begin{table}
    \centering
    \caption{Discriminator architecture and its components functions.}
    \begin{tabular}{p{0.24\columnwidth}p{0.3\columnwidth}p{0.3\columnwidth}}
        \toprule
        \textbf{Component} & \textbf{Layer Type} & \textbf{Main Function} \\
        \midrule
        Input & Concatenation of \gls{RD} map pairs & Distinguish real vs. generated data \\
        Downsampling & 3 $\times$ ($4 \times 4$ Conv + Leaky \glsxtrshort{ReLU}) & Hierarchical feature extraction \\
        Fully Connected 1 & $4 \times 4$ Conv + Leaky \glsxtrshort{ReLU} & Compresses feature representation \\
        Fully Connected 2 & $4 \times 4$ Conv + Sigmoid & Outputs authenticity score \\
        \bottomrule
    \end{tabular}
    \label{tab:discriminator_structure}
\end{table}

\subsection{Loss Functions and Training Strategy}
Different loss functions guide the training of the generator and discriminator to ensure stable adversarial learning and high-quality reconstruction.

\subsubsection{Generator Loss}
The loss function of the generator mainly comprises three key components, which are as follows.
Firstly, the \textit{L1 loss} denoted as $\mathcal{L}_{\mathrm{L1}}$, which measures the absolute difference between the generated \gls{RD} map and the \gls{HR} \gls{RD} map, ensuring structural consistency.
Secondly, the \textit{\gls{SSIM} loss} \cite{1284395}, denoted as $\mathcal{L}_\mathrm{\glsxtrshort{SSIM}}$, encourages perceptual similarity by preserving structural information.
Thirdly, the \textit{adversarial loss} \cite{10.5555/3295222.3295327} denoted as $\mathcal{L}_\mathrm{\glsxtrshort{GAN}}$, which assigns an authenticity score to the generated \gls{RD} map.
This loss function ensures that the generator learns to produce \gls{RD} maps indistinguishable from real data by maximizing the output of the discriminator, thus fooling the discriminator into classifying the generated samples as real.
The total loss of the discriminator is formulated as $\mathcal{L}_{\mathrm{G}} = \lambda_{\mathrm{L1}} \mathcal{L}_{\mathrm{L1}} + \lambda_{\mathrm{\glsxtrshort{SSIM}}} \mathcal{L}_\mathrm{\glsxtrshort{SSIM}} + \mathcal{L}_\mathrm{\glsxtrshort{GAN}}$.
The hyperparameters $\lambda_{\mathrm{L1}}$ and $\lambda_{\mathrm{\glsxtrshort{SSIM}}}$ were set to reasonable values within the range of 1 to 50, and minor variations did not significantly affect the final results.

\subsubsection{Discriminator Loss}
As proposed in \cite{10.5555/3295222.3295327}, the discriminator is trained using the Wasserstein loss with gradient penalty, mainly comprising two components.
Firstly, the \textit{Wasserstein distance}, denoted as $\mathcal{L}_{\mathrm{W}}$, measures the discrepancy between the distributions of real and generated samples.
Secondly, a \textit{\gls{GP}} loss, denoted as $\mathcal{L}_\mathrm{\glsxtrshort{GP}}$, enforces the Lipschitz constraint and regularizes the discriminator gradient norm to prevent vanishing or exploding gradients.
The total loss of the discriminator is formulated as $\mathcal{L}_{\mathrm{D}} = -\mathcal{L}_{\mathrm{W}} + \lambda_\mathrm{\glsxtrshort{GP}} \mathcal{L}_\mathrm{\glsxtrshort{GP}}$.
The gradient penalty weight is chosen as $\lambda_{\mathrm{\glsxtrshort{GP}}} = 10$, according to the configuration described in \cite{10.5555/3295222.3295327}.
\section{Results}
\label{sec:results}


\subsection{Data Set}
To train and evaluate the models, a synthetic data set is generated using the \gls{PMCW} signal model described in Section~\ref{sec:signal_model}.
The data set contains 6000 two-dimensional matrices, including 1-bit \gls{ADC} data and \gls{HR} \gls{RD} maps utilized as \gls{GT}. The \gls{HR} \gls{RD} maps are derived from unquantized \gls{ADC} data.
The signal processing takes 1-bit \gls{ADC} data as input, consisting of 3000 matrices divided into 1500 matrices with a \gls{SNR} of \SI{10}{\decibel} and 1500 matrices with an \gls{SNR} of \SI{20}{\decibel}.
For each input matrix, we generate 3000 \gls{HR} \gls{RD} maps from full-resolution \gls{ADC} data with an \gls{SNR} of \SI{50}{\decibel}. 
These \gls{HR} \gls{RD} maps serve as a reference to supervise model training.

The data set is generated using a simulated \gls{SISO} \SI{79}{\giga\hertz} automotive radar employing a \gls{MLS} of length 127.
An additional chip is added to each sequence to improve the usability of \glspl{NN}, resulting in a total length of $N = 128$.
The chip duration and bandwidth are \SI{10}{\nano\second} and \SI{100}{\mega\hertz}, respectively, with a total of $10240$ pulses being transmitted.
To improve the \gls{SNR}, we apply the accumulation approach described in \cite{7485114}.
By accumulating each set of $20$ pulses to generate a range profile, the total number of slow-time samples is reduced to $M = 512$.

\subsection{Metrics}
\label{sec:metrics}
Performance is assessed using the following metrics.
First,  the \gls{MSE} evaluates the differences between each cell of the 1-bit \gls{ADC} data and the original \gls{HR} \gls{RD} maps, which are denoted by $\mathbf{Q}^\mathrm{1b}$ and $\mathbf{Q}^\mathrm{\glsxtrshort{HR}}$, respectively, as expressed by
\begin{equation}
    \mathrm{\glsxtrshort{MSE}}\left(\mathbf{Q}^{\mathrm{1b}}, \mathbf{Q}^{\mathrm{\glsxtrshort{HR}}}\right)  = \frac{1}{NM} \sum\limits_{v=0}^{M-1}\sum\limits_{r=0}^{N-1} \Big(\left\lvert q^\mathrm{1b}_{rv}\right\rvert - \left\lvert q^\mathrm{\glsxtrshort{HR}}_{rv}\right\rvert\Big)^2,
\end{equation}
where a lower \gls{MSE} indicates better reconstruction quality.
Second, the \textit{\gls{PSL}} measures the maximum sidelobe level expressed by
\begin{equation}
     \mathrm{\glsxtrshort{PSL}}_{v}(\mathbf{Q}) = 20 \log_{10} \left(\max\limits_{r \neq \hat{r}} |q_{rv}| \right),
\end{equation}
where $v$ is the target Doppler index in \gls{RD}, and $\hat{r}$ is the range index of the main lobe peak.
Lower \gls{PSL} indicates better sidelobe suppression and, hence, better target detectability.
Third, the \textit{\gls{ISL}} measures the total sidelobes expressed by
\begin{equation}
     \mathrm{ISL}_{v}(\mathbf{Q}) = 20 \log_{10} \left(\sum\limits_{r=0, r \neq \hat{r}}^{N-1} \left\lvert q_{rv} \right\rvert^2 \right),
\end{equation}
where lower \gls{ISL} values indicate better sidelobe suppression and, similar to \gls{PSL}, improved target detectability.


\begin{table}
    \centering
    \caption{Comparison of evaluation metrics in validation scenario.} 
    \label{tab:results}
    \begin{tabular}{p{0.45\columnwidth}p{0.1\columnwidth}p{0.13\columnwidth}p{0.13\columnwidth}}
        \toprule
        \textbf{Method} & \textbf{\glsxtrshort{MSE}} & \textbf{\glsxtrshort{PSL} (\SI{}{\decibel})} & \textbf{\glsxtrshort{ISL} (\SI{}{\decibel})} \\
        \midrule
        \glsxtrshort{HR} \glsxtrshort{ADC} + \glsxtrshort{RD} & 0 & -16.1& -5.3 \\
        1-bit \glsxtrshort{ADC} + \glsxtrshort{RD} & $4.74e^{-4}$ & -18.2 & -4.6  \\
        1-bit \glsxtrshort{ADC}  + \glsxtrshort{RD} + Denoise-\glsxtrshort{NN} & $8.00e^{-6}$  & -13.5 & -5.6  \\
        1-bit \glsxtrshort{ADC} + (\glsxtrshort{RD}+Denoise)-\glsxtrshort{NN} & $1.00e^{-5}$ & -15.5 & -4.7  \\
        \bottomrule
    \end{tabular}
\end{table}

\subsection{Evaluation Based on Validation Scenario}
As illustrated in Table~\ref{tab:generator_structure}, the generator employs an \gls{E2E} approach to transform 1-bit \gls{ADC} data into \gls{HR} \gls{RD} maps, implicitly fulfilling two tasks jointly: \gls{RD} processing and denoising.
This approach is hereinafter referred to as \textit{1-bit \gls{ADC} + (\gls{RD}+Denoise)-\gls{NN}}.
In contrast, using the hybrid approach, the generator consists only of the denoising module, further referred to as \textit{1-bit \gls{ADC} + \gls{RD} + Denoise-\gls{NN}}. 
The conversion of \gls{ADC} data to \gls{RD} maps, as detailed in Section~\ref{sec:signal_model}, occurs before the application of the denoising \gls{NN}.

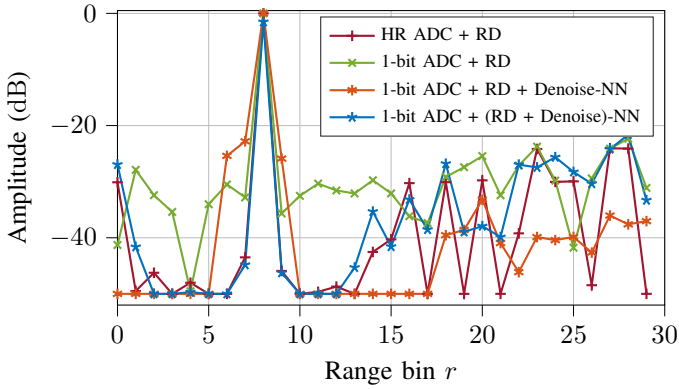
\begin{figure}
    \centering
\begin{tikzpicture}

\definecolor{mycolor1}{rgb}{0.00000,0.44700,0.74100}%
\definecolor{mycolor2}{rgb}{0.85000,0.32500,0.09800}%
\definecolor{mycolor3}{rgb}{0.92900,0.69400,0.12500}%
\definecolor{mycolor4}{rgb}{0.49400,0.18400,0.55600}%
\definecolor{mycolor5}{rgb}{0.46600,0.67400,0.18800}%
\definecolor{mycolor6}{rgb}{0.30100,0.74500,0.93300}%
\definecolor{mycolor7}{rgb}{0.63500,0.07800,0.18400}%

\begin{axis}[
height=5.5cm,
width=\columnwidth,
tick align=outside,
tick pos=left,
xlabel={Range bin $r$},
xmin=0, xmax=30.,
xtick style={color=black},
ylabel={Amplitude (\SI{}{\decibel}) },
ymin=-52, ymax=0.5,
ytick style={color=black},
ymajorgrids,
xmajorgrids,
legend style={legend cell align=left, draw=white!15!black, font=\scriptsize},
]
\addplot [thick, mycolor7, mark=+]
table {%
0 -30.1102714538574
1 -49.488639831543
2 -46.2375297546387
3 -50
4 -48.0190277099609
5 -50
6 -50
7 -43.49462890625
8 0
9 -45.9154815673828
10 -50
11 -49.6382331848145
12 -48.7093772888184
13 -50
14 -42.5338325500488
15 -40.2872009277344
16 -30.2637939453125
17 -50
18 -30.0864772796631
19 -50
20 -29.7678604125977
21 -50
22 -39.2140960693359
23 -24.1696281433105
24 -30.0394325256348
25 -29.94358253479
26 -48.4597091674805
27 -24.0426235198975
28 -24.1091060638428
29 -50
};
\addlegendentry{\glsxtrshort{HR} \glsxtrshort{ADC} + \gls{RD}}

\addplot [thick, mycolor5, mark=x]
table {%
0 -41.2893867492676
1 -27.895357131958
2 -32.415153503418
3 -35.3918724060059
4 -49.5344886779785
5 -34.0239639282227
6 -30.5012741088867
7 -32.8074417114258
8 0
9 -35.6643257141113
10 -32.5305709838867
11 -30.3435878753662
12 -31.5712013244629
13 -32.1033248901367
14 -29.7474746704102
15 -32.0907936096191
16 -36.1662979125977
17 -37.3716125488281
18 -29.1693572998047
19 -27.4178333282471
20 -25.4411716461182
21 -32.4280586242676
22 -27.1318225860596
23 -23.7562389373779
24 -29.8096446990967
25 -41.772834777832
26 -29.4091777801514
27 -24.1148452758789
28 -22.3214893341064
29 -31.1125106811523
};
\addlegendentry{1-bit \glsxtrshort{ADC} + \glsxtrshort{RD}}

\addplot [thick, mycolor2, mark=asterisk]
table {%
0 -50
1 -50
2 -50
3 -50
4 -50
5 -50
6 -25.4017219543457
7 -22.8041095733643
8 0
9 -25.888126373291
10 -50
11 -50
12 -50
13 -50
14 -50
15 -50
16 -50
17 -50
18 -39.5412139892578
19 -38.6160011291504
20 -33.2152252197266
21 -40.9496459960938
22 -46.1314849853516
23 -39.878734588623
24 -40.4236221313477
25 -39.8570709228516
26 -42.6443672180176
27 -36.0396423339844
28 -37.6086730957031
29 -37.03662109375
};
\addlegendentry{1-bit \glsxtrshort{ADC} + \glsxtrshort{RD} + Denoise-\glsxtrshort{NN}}

\addplot [thick, mycolor1, mark=star]
table {%
0 -26.9477252960205
1 -41.6301727294922
2 -50
3 -50
4 -49.7044448852539
5 -50
6 -50
7 -44.8251686096191
8 -1.52560758590698
9 -46.2468032836914
10 -50
11 -50
12 -50
13 -45.3084564208984
14 -35.3305358886719
15 -41.5828514099121
16 -33.1782073974609
17 -38.5466613769531
18 -26.7957992553711
19 -38.9596099853516
20 -37.9217834472656
21 -39.8687629699707
22 -26.9652519226074
23 -27.4574775695801
24 -25.6364784240723
25 -28.2699699401855
26 -30.3675155639648
27 -24.1935749053955
28 -21.6100521087646
29 -33.3113059997559
};
\addlegendentry{1-bit \glsxtrshort{ADC} + (\glsxtrshort{RD} + Denoise)-\glsxtrshort{NN}}

\end{axis}

\end{tikzpicture}
    \caption{Slices of range-Doppler map for a validation scenario with a single target. The curves for the different approaches have been normalized to peak power.} 
    \label{fig:validation_scenario}
\end{figure}

As shown in Table~\ref{tab:results}, each method exhibits distinct strengths in terms of \gls{MSE}, \gls{PSL}, and \gls{ISL}. 
As detailed in Section~\ref{sec:metrics}, the  \gls{MSE} is determined considering all cells within the final \gls{RD} maps, whereas \gls{PSL} and \gls{ISL} are explicitly computed for the velocity bin depicted in Fig.~\ref{fig:validation_scenario}.
Thus, the \gls{MSE} serves as a more comprehensive metric, while \gls{PSL} and \gls{ISL} are localized metrics tailored to individual target velocity bins.
The conventional \textit{1-bit \gls{ADC} + \gls{RD}} approach achieves the best \gls{PSL} in the given scenario; however, its total \gls{MSE} exceeds that of the \gls{NN}-based methods, indicating that while it lowers peak sidelobes, the noise level is elevated due to quantization effects.
Moreover, its \gls{ISL} is greater than for the other methods.
When integrating 1-bit \gls{ADC} and \gls{RD} processing with a denoising-\glsxtrshort{NN} in a hybrid approach (\textit{1-bit \gls{ADC} + \gls{RD} + Denoise-\gls{NN}}), there is a reduction in the \gls{MSE} and an improvement in the \gls{ISL}. 
However, this combination results in a smoothing effect during suppression of the peak sidelobes. 
While background noise is reduced, the relative amplitudes of certain sidelobes increase, causing an increase in the \gls{PSL}.
The loss function mainly focuses on minimizing pixel errors and lacks explicit conditions regarding peak sidelobes so that certain sidelobe levels may rise.
Future research might consider incorporating the \gls{PSL} and \gls{ISL} metrics within the loss function.
Integrating \gls{RD} processing and denoising into an \gls{E2E} \gls{NN} (\textit{1-bit \gls{ADC} + (\gls{RD}+Denoise)-\gls{NN}}) leads to a more balanced performance in all metrics. 
The \glspl{MSE} are calculated relative to the \textit{\gls{HR} \gls{ADC} + \gls{RD}} method, where the unquantized \gls{ADC} data undergo conventional \gls{RD} processing.
This shows that the \gls{E2E} generator module can efficiently extract range and Doppler information from the \gls{ADC} data, illustrating the adaptability of the developed \gls{GAN}.
It can be inferred that a denoising \gls{NN} on quantized \gls{ADC} data can mitigate quantization impacts and improve target detectability. 

\subsection{Computational Complexity and Resource Consumption}
The hybrid approach slightly reduces the \gls{FLOPS}, and compared with the \gls{E2E} approach, its inference speed is greatly improved, decreasing from \SI{3.1}{\second} to \SI{1.7}{\second}. 
Experimental results based on \textit{Nvidia GeForce GTX 1080 Ti} show that the hybrid approach requires an average of \SI{2800}{\second} per epoch during training, while the \gls{E2E} approach takes about \SI{6000}{\second} per epoch. 
These results demonstrate that the hybrid approach accelerates inference and significantly shortens training time, thereby enhancing its appeal for real-time applications.
\section{Conclusion and Outlook}
\label{sec:conclusion}

Although the \gls{E2E} generator slightly improves reconstruction quality, the experimental results indicate that conventional \gls{RD} processing can achieve comparable performance when combined with a subsequent denoising \gls{NN}. 
In addition, conventional processing benefits from reduced training time and lower computational complexity.
Given the constraints on real-time processing and resource limitations, the conventional approach is more attractive for practical applications.


To prove effectiveness under various conditions, networks should be trained using various scenarios, including more targets, target \glspl{RCS}, and \glspl{SNR}. 
Furthermore, the performance of the model should be assessed by applying it to actual radar systems.

\bibliographystyle{IEEEtran}
\bibliography{bib}

\vspace{12pt}

\end{document}